\documentclass[apj]{emulateapj}
\slugcomment{Accepted for publication in The Astrophysical Journal Letters}

\shorttitle{On the Location of the $\gamma$-ray Outburst Emission in \object{AO~0235+164}}
\shortauthors{Agudo et al.}

\begin{document}

\title{On the Location of the $\gamma$-ray Outburst Emission in the BL~Lacertae Object \object{AO~0235+164} through Observations across the Electromagnetic Spectrum}

\author{Iv\'{a}n Agudo\altaffilmark{1,2},
            Alan P.~Marscher\altaffilmark{2},
            Svetlana G.~Jorstad\altaffilmark{2,3},
            Valeri M.~Larionov\altaffilmark{3,4},
            Jos\'{e} L.~G\'omez\altaffilmark{1},
            Anne L\"{a}hteenm\"{a}ki\altaffilmark{5},
            Paul S. Smith\altaffilmark{6},
            Kari Nilsson\altaffilmark{7},
            Anthony C.~S.~Readhead\altaffilmark{8},
            Margo F.~Aller\altaffilmark{9},
            Jochen Heidt\altaffilmark{10}, 
            Mark Gurwell\altaffilmark{11},
            Clemens Thum\altaffilmark{12},
            Ann E.~Wehrle\altaffilmark{13}, 
            Maria G.~Nikolashvili\altaffilmark{14},
            Hugh D.~Aller\altaffilmark{9},    
            Erika Ben\'{i}tez\altaffilmark{15},
            Dmitriy A.~Blinov\altaffilmark{3,4},
            Vladimir A.~Hagen-Thorn\altaffilmark{3,4}, 
            David Hiriart\altaffilmark{16},
            Buell T.~Jannuzi\altaffilmark{17},
            Manasvita Joshi\altaffilmark{2},
            Givi N.~Kimeridze\altaffilmark{14},
            Omar M.~Kurtanidze\altaffilmark{14},
            Sofia O.~Kurtanidze\altaffilmark{14},
            Elina Lindfors\altaffilmark{18},
            Sol N. Molina\altaffilmark{1}, 
            Daria A.~Morozova\altaffilmark{3},
            Elina Nieppola\altaffilmark{5,7},
            Alice R.~Olmstead\altaffilmark{2},
            Riho Reinthal\altaffilmark{18},
            Mar Roca--Sogorb\altaffilmark{1},
            Gary D.~Schmidt\altaffilmark{19},
            Lorand A.~Sigua\altaffilmark{14},
            Aimo Sillanp\"a\"a\altaffilmark{18},
            Leo Takalo\altaffilmark{18},
            Brian Taylor\altaffilmark{2,20},
            Merja Tornikoski\altaffilmark{5},
            Ivan S.~Troitsky\altaffilmark{3},
            Alma C.~Zook\altaffilmark{21},
            Helmut Wiesemeyer\altaffilmark{22}
           }


\altaffiltext{1}{Instituto de Astrof\'{i}sica de Andaluc\'{i}a, CSIC, Apartado 3004, 18080, Granada, Spain; \email{iagudo@iaa.es}}

\altaffiltext{2}{Institute for Astrophysical Research, Boston University, 725 Commonwealth Avenue, Boston, MA 02215, USA}

\altaffiltext{3}{Astronomical Institute, St. Petersburg State University, Universitetskij Pr. 28, Petrodvorets, 198504 St. Petersburg, Russia}

\altaffiltext{4}{Isaac Newton Institute of Chile, St. Petersburg Branch, St. Petersburg, Russia}

\altaffiltext{5}{Aalto University Mets\"{a}hovi Radio Observatory, Mets\"{a}hovintie 114, FIN-02540 Kylm\"{a}l\"{a}, Finland}

\altaffiltext{6}{Steward Observatory, University of Arizona, Tucson, AZ 85721-0065, USA}

\altaffiltext{7}{Finnish Centre for Astronomy with ESO (FINCA), University of Turku, V\"ais\"al\"antie 20, FIN-21500 Piikki\"o, Finland}

\altaffiltext{8}{Cahill Center for Astronomy and Astrophysics, California Institute of Technology, Mail Code 222, Pasadena, CA 91125, USA}

\altaffiltext{9}{Department of Astronomy, University of Michigan, 817 Dennison Building, Ann Arbor, MI 48 109, USA}

\altaffiltext{10}{ZAH, Landessternwarte Heidelberg, K\"{o}nigstuhl, 69117 Heidelberg, Germany}

\altaffiltext{11}{Harvard--Smithsonian Center for Astrophysics, 60 Garden St., Cambridge, MA 02138, USA}

\altaffiltext{12}{Institut de Radio Astronomie Millim\'{e}trique, 300 Rue de la Piscine, 38406 St. Martin d'H\`{e}res, France}

\altaffiltext{13}{Space Science Institute, Boulder, CO 80301, USA}

\altaffiltext{14}{Abastumani Observatory, Mt. Kanobili, 0301 Abastumani, Georgia}

\altaffiltext{15}{Instituto de Astronom\'{i}a, Universidad Nacional Aut\'{o}noma de M\'{e}xico, 04510 M\'{e}xico D.~F., M\'{e}xico}

\altaffiltext{16}{Instituto de Astronom\'{i}a, Universidad Nacional Aut\'{o}noma de M\'{e}xico, 2280 Ensenada, M\'{e}xico}

\altaffiltext{17}{National Optical Astronomy Observatory, KPNO, Tucson, AZ 85726, USA}

\altaffiltext{18}{Tuorla Observatory, University of Turku, V\"ais\"al\"antie 20, FIN-21500 Piikki\"o, Finland}

\altaffiltext{19}{National Science Foundation, 4201 Wilson Blvd., Arlington, VA 22230, USA}

\altaffiltext{20}{Lowell Observatory, Flagstaff, AZ 86001, USA}

\altaffiltext{21}{Department of Physics and Astronomy, Pomona College, Claremont, CA 91711,USA}

\altaffiltext{22}{Instituto de Radio Astronom\'{i}a Milim\'{e}trica, Avenida Divina Pastora, 7, Local 20, E--18012 Granada, Spain}

\begin{abstract}
We present observations of a major outburst at centimeter, millimeter, optical, X-ray, and $\gamma$-ray wavelengths of the BL~Lacertae object \object{AO~0235+164}. 
We analyze the timing of multi-waveband variations in the flux and linear polarization, as well as changes in Very Long Baseline Array (VLBA) images at $\lambda=7$\,mm with $\sim0.15$ milliarcsecond resolution. 
The association of the events at different wavebands is confirmed at high statistical significance by probability arguments and Monte-Carlo simulations. 
A series of sharp peaks in optical linear polarization, as well as a pronounced maximum in the 7~mm polarization of a superluminal jet knot, indicate rapid fluctuations in the degree of ordering of the magnetic field.  
These results lead us to conclude that the outburst occurred in the jet both in the quasi-stationary ``core'' and in the superluminal knot, both parsecs downstream of the supermassive black hole. 
We interpret the outburst as a consequence of the propagation of a disturbance, elongated along the line of sight by light-travel time delays, that passes through a standing recollimation shock in the core and propagates down the jet to create the superluminal knot. 
The multi-wavelength light curves vary together on long time-scales (months/years), but the correspondence is poorer on shorter time-scales. 
This, as well as the variability of the polarization and the dual location of the outburst, agrees with the expectations of a multi-zone emission model in which turbulence plays a major role in modulating the synchrotron and inverse Compton fluxes.   
\end{abstract}


\section{Introduction}
\label{intr}

Our understanding of the processes leading to the generation of $\gamma$-ray emission from blazars, the most extreme active galactic nuclei, depends on where those $\gamma$-rays originate. 
This is currently the subject of considerable debate \citep[e.g.,][]{Marscher:2010p14415,Tavecchio:2010p14858,Agudo:2011p14707}.
Two main locations of the site of $\gamma$-ray emission in blazars have been proposed.
The first is close ($\lesssim0.1{\rm{-}}1$\,pc) to the supermassive black hole (BH), which can readily explain short time-scales of variability of a few hours reported in some $\gamma$-ray observations of blazars \citep{Ackermann:2010p13506, Foschini:2010p13037,Foschini:2011p14843, Tavecchio:2010p14858}.
Although short time-scales of variability limit the size of the emission region and does not necessarily imply short distances to the BH, this scenario has the advantage that optical-UV photons from the broad emission-line region are available for scattering to $\gamma$-ray energies by highly relativistic electrons in the jet.

However, locating the $\gamma$-ray emission region so close to the BH contradicts the increasingly large number of time coincidences of radio-millimeter and $\gamma$-ray events \citep{Jorstad:2010p11830,Marscher:2010p11374,Agudo:2011p14707}.
This problem is overcome if the $\gamma$-rays are emitted from a region much farther ($>>1$\,pc) from the BH, beyond the ``core'' where the jet starts to be visible at millimeter wavelengths with very long baseline interferometry (VLBI).
Supporting this second scenario, \citet{Agudo:2011p14707} unambiguously locate the region of $\gamma$-ray flares $>14$\,pc from the BH in the jet of \object{OJ287} through correlation of millimeter-wave with $\gamma$-ray light curves and direct ultrahigh-resolution 7\,mm imaging with the VLBA.
Similar results are obtained by \citet{Marscher:2010p11374} and \citet{Jorstad:2010p11830} for \object{PKS 1510$-$089} and \object{3C~454.3}, respectively.
\citet{Marscher:2010p14415} have recently proposed a model that reconciles both $\gamma$-ray flare emitting regions located at $>>1$\,pc from the BH and intraday $\gamma$-ray variability through a model involving turbulent plasma flowing through standing shocks in the core with a volume filling factor that decreases with energy and therefore frequency.

This paper investigates the location and properties of a radio to $\gamma$-ray outburst in the BL~Lacertae object \object{AO~0235+164} (\object{0235+164} hereafter, $z=0.94$). 
Historically, the blazar has exhibited extreme variability (by over an order of magnitude on time-scales $<1$\,year) across all spectral ranges, including X-ray \cite[e.g.,][]{Raiteri:2009p9413} and $\gamma$-ray \citep[e.g.,][]{Abdo:2010p11947}. 
The most prominent radio outbursts are accompanied by optical counterparts \citep{Ledden:1976p15696,Balonek:1980p15693,Raiteri:2008p13061}, although no correlation with variability at other wavebands has been reported previously.

On sub-milliarcsecond scales, \object{0235+164} is extremely compact ($\lesssim0.5$\,mas) at millimeter wavelengths \citep[e.g.,][]{Piner:2006p12065}, and shows extreme superluminal apparent speeds \citep[$\beta_{\rm{app}}^{\rm{max}}=46.5\pm8.0\,c$,][]{Jorstad:2001p5655}\footnote{We adopt the standard $\Lambda$CDM cosmology, with $H_0$=71 km s$^{-1}$ Mpc$^{-1}$, $\Omega_M=0.27$, and $\Omega_\Lambda=0.73$, so that 1\,mas corresponds to a projected linear distance of 7.9\,pc.}.
This speed, and the variability Doppler factor $\delta_{var}=24.0$ of \object{0235+164} \citep{Hovatta:2009p3860}, sets a maximum jet viewing angle $\theta=\arctan[2\beta_{\rm{app}}^{\rm{max}}/({\beta_{\rm{app}}^{\rm{max}}}^{2}+\delta_{var}^{2}-1)]\lesssim2^\circ\kern-.35em .4$.
Although the large $\beta_{\rm{app}}^{\rm{max}}$ value might lead to underestimation of $\theta$ for non-cylindric jets \citep{GopalKrishna:2006p15835}, the unusual compactness of \object{0235+164} and its extreme flaring activity is consistent with low $\theta$ values as estimated above \citep[see also][]{Hovatta:2009p3860}.
Superluminal knots have been observed at position angles ranging from $5^{\circ}$ to $-55^{\circ}$ \citep{Jorstad:2001p5655,Piner:2006p12065}, consistent with a broad ($\alpha_{\rm{app}}\approx60^{\circ}$) projected jet opening-angle.
Both $\theta$ and $\alpha_{\rm{app}}$ constrain the maximum intrinsic jet half--opening--angle $\alpha_{\rm{int}}/2=(\alpha_{\rm{app}}\sin\theta)/2\lesssim1^\circ\kern-.35em .25$.

\section{Observations}
\label{obs}

\begin{figure*}
   \centering
   \includegraphics[clip,width=16.cm]{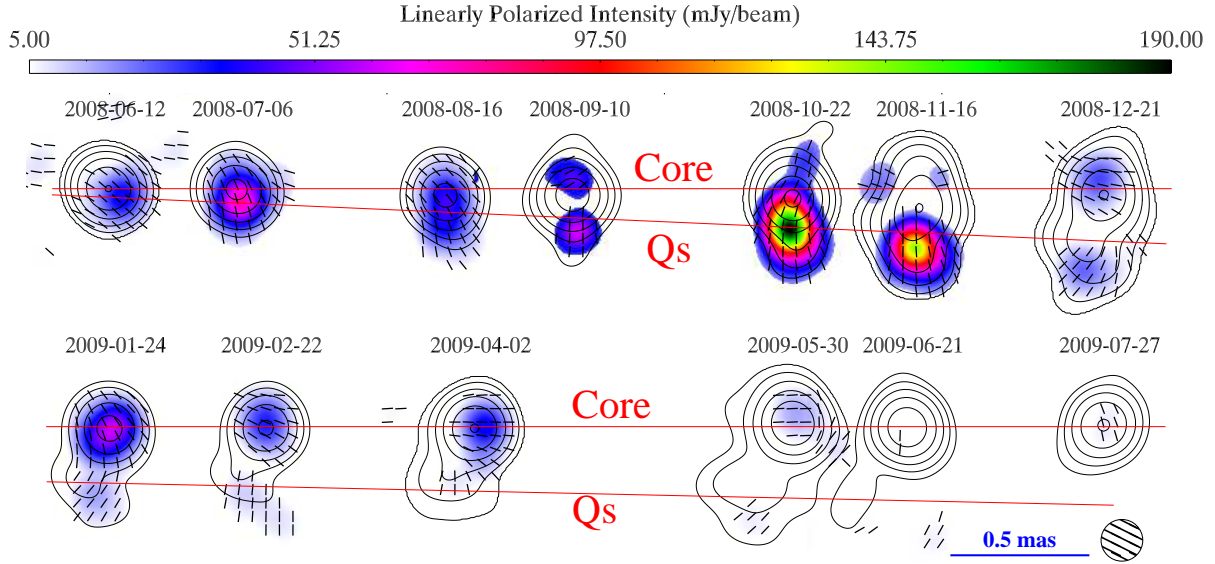}
   \caption{Sequence of 7\,mm VLBA images of 0235+164 convolved with a $\rm{FWHM}=0.15$\,mas circular Gaussian beam. Images in our program before 2008-06-12 and after 2009-07-27, containing only a single emission feature (i.e., the core), are not displayed. Contour levels represent total intensity (levels in factors of 2 from $0.4$ to  $51.2$\,\% plus $90.0$\,\% of peak$=4.93$\,Jy/beam), color scale indicates polarized intensity, and superimposed sticks show the orientation of $\chi$.}
   \label{maps}
\end{figure*}

Our photo-polarimetric monitoring observations of \object{0235+164} (Figs.~\ref{maps}-\ref{pol}) include (1) 7\,mm images with the VLBA from the Boston University monthly blazar-monitoring program\footnote{\tt http://web.bu.edu/blazars/VLBAproject.html}, (2) 3\,mm observations with the IRAM 30m Telescope, and (3) optical measurements from the following telescopes: Calar Alto (2.2m Telescope, observations under the MAPCAT\footnote{\tt http://www.iaa.es/$\sim$iagudo/research/MAPCAT} program), Steward Observatory (2.3 and 1.54m Telescopes\footnote{{\tt http://james.as.arizona.edu/$\sim$psmith/Fermi}}), Lowell Observatory (1.83m Perkins Telescope), San Pedro M\'{a}rtir Observatory (0.84m Telescope), Crimean Astrophysical Observatory (0.7m Telescope), and St. Petersburg State University (0.4m Telescope). 
Our total flux light curves (Fig.~\ref{tflux}) include data from the \emph{Fermi}-LAT $\gamma$-ray (0.1--200\,GeV) and \emph{Swift}-XRT X-ray (2.4--10\,keV) observatories, available from the archives of these missions, and RXTE at 2.4--10\,keV.
Optical $R$-band fluxes come from the Tuorla Blazar Monitoring Program\footnote{{\tt http://users.utu.fi/kani/1m}}, the Yale University SMARTS program\footnote{{\tt http://www.astro.yale.edu/smarts/glast}}, and Maria Mitchell and Abastumani Observatories.
Longer wavelength light-curves were acquired from the Submillimeter Array (SMA) at 850\,$\mu$m and 1\,mm, the IRAM 30m Telescope at 1\,mm, the Mets\"{a}hovi 14m Telescope at 8\,mm, and both the Owens Valley Radio Observatory (OVRO) 40m Telescope \emph{Fermi} Blazar Monitoring Program\footnote{\tt http://www.astro.caltech.edu/ovroblazars/} and University of Michigan Radio Astronomy Observatory\footnote{\tt http://www.astro.lsa.umich.edu/obs/radiotel/} (UMRAO) 26m Telescope at 2\,cm.

We followed data reduction procedures described in previous studies: VLBA: \citet{Jorstad:2005p264}; optical polarimetric data: \citet{Jorstad:2010p11830}; IRAM data: \citet{Agudo:2006p203, Agudo:2010p12104}; SMA: \citet{Gurwell:2007p12057}; Mets\"{a}hovi: \citet{1998A&AS..132..305T}; OVRO: \citet{Richards:2010p14140}; UMRAO: \citet{Aller:1985p6715}; \emph{Swift}: \citet{Jorstad:2010p11830}; \emph{RXTE:} \citet{Marscher:2010p11374}; and \emph{Fermi}-LAT: \citet{Marscher:2010p11374, Agudo:2011p14707}. 
For both \emph{Swift} and RXTE data, only the 2.4--10\,keV band is included in the X-ray light curve.
In this energy band, where the intrinsic spectral curvature and photoelectric absorption at $z=0.524$ discussed by \citet[][and references therein]{Raiteri:2006p13056} are not important, a single absorbed power law (with Galactic absorption corresponding to a neutral hydrogen column density of $0.9\times10^{21}$\,cm$^2$) fit the data adequately.
To process the 0.1--200~GeV \emph{Fermi}-LAT data, we used 2-day binning and a single power-law spectral fit with photon index held fixed at 2.14 \citep{Abdo:2010p12082}. 
This produces essentially the same light curve as the broken-power-law spectral model of \citet{Abdo:2010p14081}.

\section{Analysis of Observational Results}

\subsection{Major Millimeter Flare in 2008 Related to a New Superluminal Knot}
\label{ejection}

Our 7\,mm VLBA maps of \object{0235+164} (Fig.~\ref{maps}) reveal compact total intensity structure that is well fitted by one or two circular Gaussian components at most of the observing epochs, where the core generally corresponds to the brighter one.
Between mid-2008 and mid-2009, the images also contain a second emission region (Qs, the brightest jet feature detected thus far in this object) that propagates at a superluminal apparent speed $<\beta_{\rm{app}}>=(12.6\pm1.2)\,c$.
The position angle of Qs$\in[\sim150^{\circ},\sim180^{\circ}]$ contrasts with previously reported VLBI position angles by up to $\sim180^{\circ}$, which is consistent with a sudden change of the jet ejection angle amplified by projection effects.

\begin{figure*}
   \centering
   \includegraphics[clip,width= 16.cm]{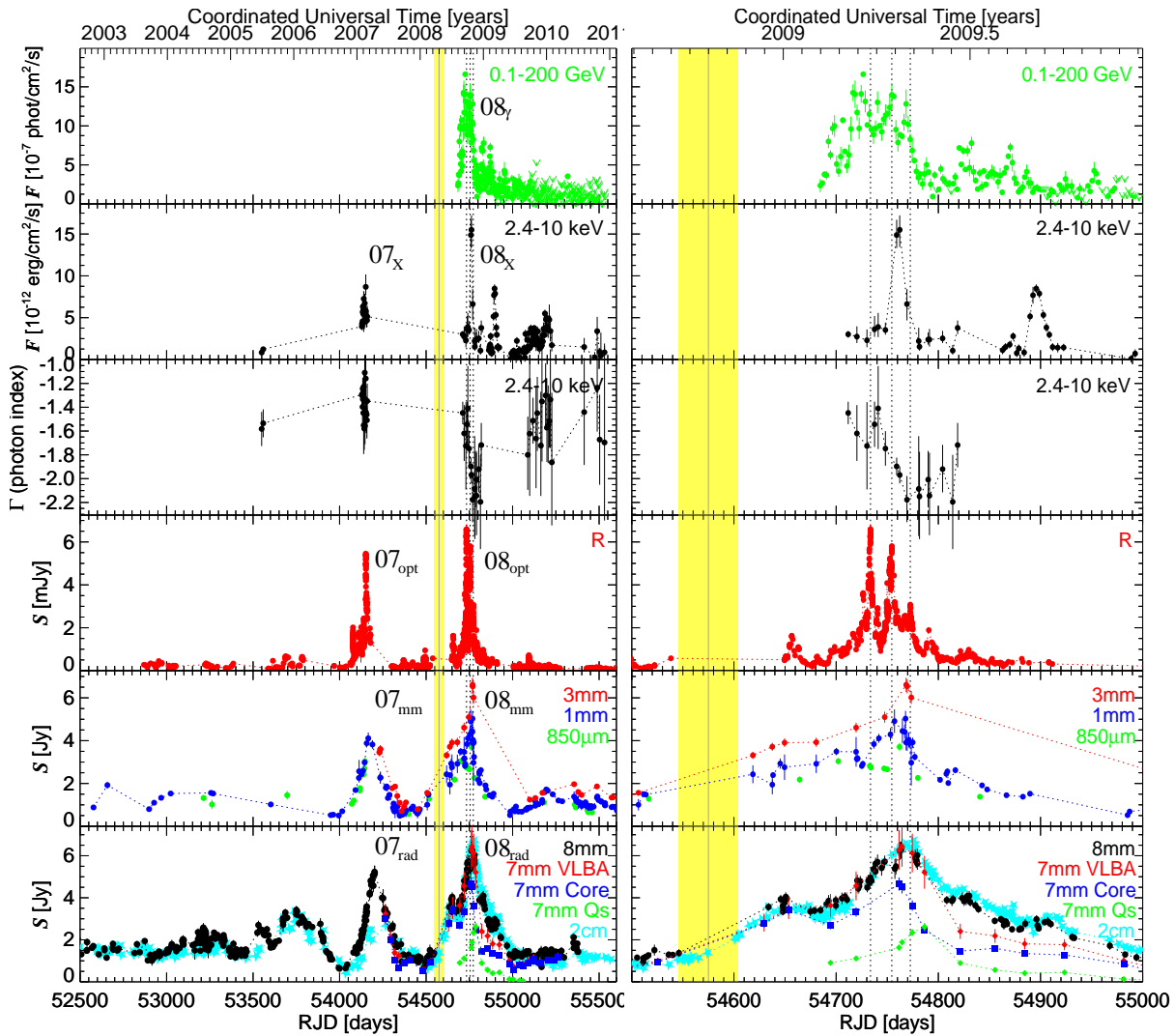}
   \caption{\emph{Left:} Light curves of 0235+164 from $\gamma$-ray to millimeter wavelengths. 
   X-ray photon index evolution from the \emph{Swift}-XRT data is also plotted. 
   Vertical dotted lines mark the three most prominent ${\rm{08}}_{\rm{opt}}$ optical peaks. The yellow area represents the time of ejection of feature Qs within its uncertainty. RJD = Julian Date~$-$~2400000.0. \emph{Right:} Same as left panel for RJD$\in[54500,55000]$.}
  \label{tflux}
\end{figure*}

If the motion of Qs is ballistic, it was coincident with the core in $2008.30\pm0.08$, near the start of an extreme millimeter-wave outburst (labeled ${\rm{08}}_{\rm{mm}}$ in Fig.~\ref{tflux}) that peaked on 2008 October 10 with a flux density $\sim6.5$\,Jy at 3\,mm.
Figure~\ref{tflux} shows that radio and millimeter-wave outbursts in 2008 (${\rm{08}}_{\rm{rad}}$ and ${\rm{08}}_{\rm{mm}}$) contain contributions from both the core and Qs, whose fluxes reached maximum on 2008 October 20 and November 16, respectively.
Their contemporaneous co-evolution suggests that the disturbance responsible for the ejection of Qs extended from the location of the core to Qs in the frame of the observer, which could have resulted from light-travel delays \citep[e.g.,][]{1997ApJ...482L..33G,Agudo:2001p460}.
Qs is the brighter 7-mm superluminal knot ever seen in \object{0235+164}, and flares ${\rm{08}}_{\rm{rad}}$ and ${\rm{08}}_{\rm{mm}}$ are the only outbursts that occurred after the ejection of Qs. 
The rarity of such events strongly implies that they are physically related.

The jet half--opening--angle of \object{0235+164} ($\alpha_{\rm{int}}/2\lesssim1^\circ\kern-.35em .25$) and the average FWHM of the core measured from our 31 VLBA observing epochs in [2007,2010] ($\langle{\rm{FWHM}_{\rm{core}}}\rangle=(0.054\pm0.018)$\,mas), constrain the 7\,mm core to be at $d_{\rm{core}}=1.8 \langle{\rm{FWHM}_{\rm{core}}\rangle/\tan\alpha_{\rm{int}}}\gtrsim12$\,pc from the vertex of the jet cone.

\subsection{Contemporaneous Flares from $\gamma$-ray to Radio Wavelengths}
\label{corr}
\begin{figure*}
   \centering
   \includegraphics[clip,width= 16.cm]{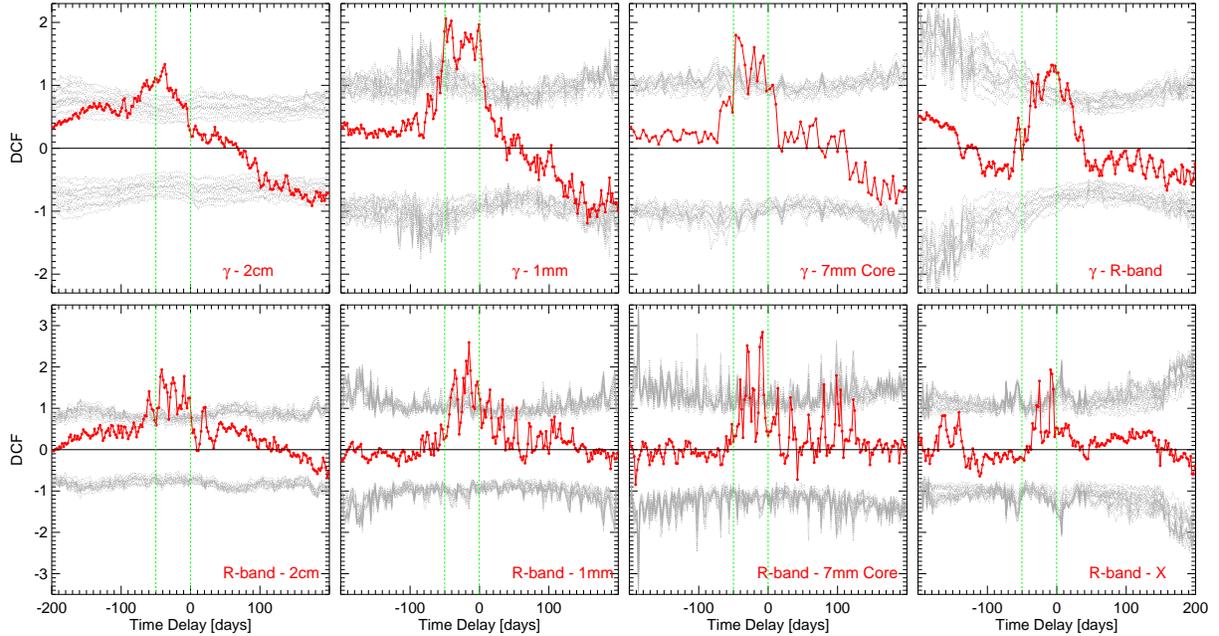}
   \caption{Grid of DCF \citep{Edelson:1988p12129} of labeled light-curve pairs during the maximum time period ${\rm{RJD}}=[52200,55600]$. \emph{Top} row of panels show DFC with $\gamma$-ray light-curve, whereas \emph{bottom} row show DFV with $R$-band light-curve. Grey dotted curves at positive (negative) DCF values symbolize 99.7\,\% confidence limits for correlation against the null hypothesis of stochastic variability. Each of these curves corresponds to a Monte-Carlo simulation of 5000 $\gamma$-ray and ${\lambda{2}}$ (where ${\lambda{2}}$ denotes the lower energy spectral range) light curves characterized by the same mean and standard deviation as those of the observed light curves, and by power--law power spectral densities (PSD$\propto1/f^{a}$) with $a_{\gamma}\in\{1.0,1.5,2.0\}$ and $a_{\lambda{2}}\in\{1.0,1.5,2.0,2.5,3.0\}$. Green dashed lines at $0$ and $-50$ day time--lags are drawn for reference.}
   \label{dcf1}
\end{figure*}

Figure~\ref{tflux} reveals that the ${\rm{08}}_{\rm{rad}}$ and ${\rm{08}}_{\rm{mm}}$ flares were accompanied by sharp optical, X--ray, and $\gamma$-ray counterparts (${\rm{08}}_{\rm{opt}}$, ${\rm{08}}_{\rm{X}}$, and  ${\rm{08}}_{\gamma}$ flares, respectively). 
Our formal light-curve correlation analysis (Fig.~\ref{dcf1}) --performed following \citet{Agudo:2011p14707}-- confirms the association of $\gamma$-ray variability with that at 2\,cm, 8\,mm, 1\,mm, and optical wavelengths at $>99.7$\,\% confidence.
The flux evolution of the VLBI core is also correlated with the $\gamma$-ray light curve at $>99.7$\,\% confidence.
Moreover, the evolution of the degree of optical linear polarization ($p_{\rm{opt}}$) and X-ray light curve are also correlated with the optical $R$-band, 1\,mm, and 2\,cm light curves at $>99.7$\,\% confidence (Fig.~\ref{dcf2}), further indicating that the extreme flaring activity revealed by our light curves is physically related at all wavebands from radio to $\gamma$-rays.

There is, however, no common pattern to the discrete correlation function (DCF) at all spectral ranges. 
This implies that, although there is correlation on long time-scales (years), on short time-scales ($\lesssim2$~months) the variability pattern does not correspond as closely.
This is the result of the intrinsic variability pattern rather than the irregular time sampling at some spectral ranges.

The sharp systematic peaks in the DCFs involving the $R$-band light curve aids in the identification of relative time delays across wavebands, as measured with regard to the first sharp $\rm{DCF}_{\rm{R},\lambda2}$ peak.
In this way, we find that, relative to the $R$-band maximum, the peaks of the $\lambda2=2$\,cm, 8\,mm, 1\,mm, and 7\,mm core flares in 2008 are delayed by $\sim60$,  $\sim45$,  $\sim40$, and $\sim40$ days, respectively, whereas the X-ray delay is $\sim25$ days. 
Only the $\gamma$-ray variations lead those at $R$-band, by $\sim10$\,days according to the peak in the DCF seen in Figure~\ref{dcf1}.

\begin{figure*}
   \centering
   \includegraphics[clip,width= 16.cm]{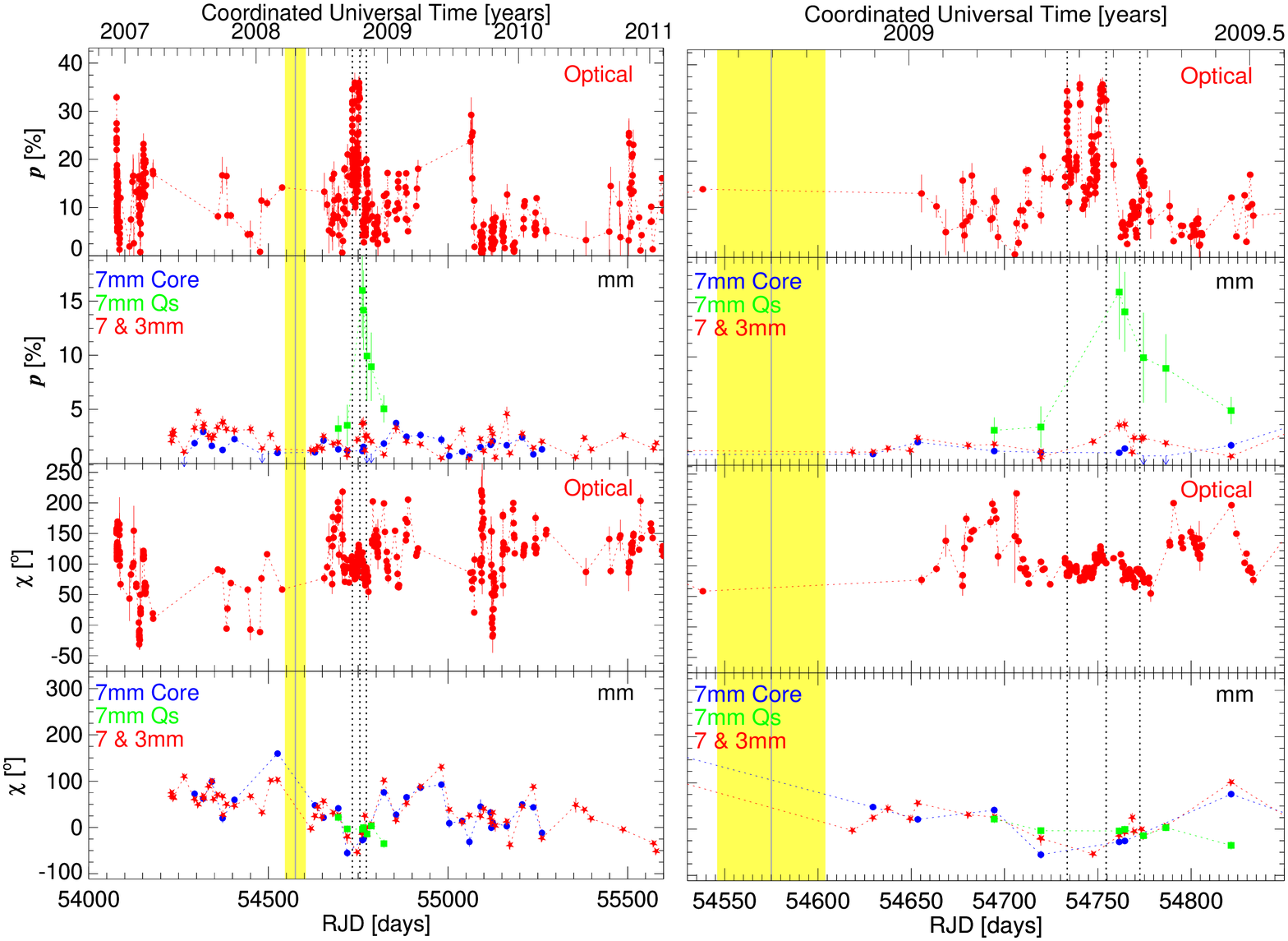}
   \caption{\emph{Left:} Long term optical and millimeter-wave linear polarization evolution of 0235+164 in the RJD=[54000,55600] range. \emph{Right:} Same as left panel for RJD$\in[54530,54850]$.}
   \label{pol}
\end{figure*}

\subsection{Correlated Variability of Linear Polarization}
\label{polvar}
Figure~\ref{pol} reveals extremely high, variable optical polarization, $p_{\rm{opt}}\gtrsim30$\,\%, during the sharp ${\rm{08}}_{\rm{opt}}$ optical peaks \citep[see also][for the 2006--2007 outburst]{HagenThorn:2008p13118}.
Whereas the integrated millimeter-wave degree of linear polarization ($p_{\rm{mm}}$) and that of the 7~mm core remain at moderate levels $\lesssim5$\,\%, the polarization of Qs ($p_{\rm{mm,Qs}}$) peaks at the high value of $\sim16$\,\% close to the time of the second sharp optical sub-flare.
The coincidence of this sharp maximum of $p_{\rm{mm,Qs}}$ in the brightest superluminal feature ever detected in \object{0235+164} with the (1) high optical flux and polarization, (2) flares across the other spectral regimes, and (3) flare in the 7-mm VLBI core, implies that the ejection and propagation of Qs in \object{0235+164}'s jet is physically tied to the total flux and polarization variations from radio to $\gamma$-rays.

On long time-scales (years), the linear polarization angle at both optical ($\chi_{\rm{opt}}$) and millimeter ($\chi_{\rm{mm}}$ and $\chi_{\rm{mm}}^{\rm{core}}$) wavelengths varies wildly, without a preferred orientation or systematic common trend. 
However, during flare ${\rm{08}}_{\rm{opt}}$, $\chi_{\rm{opt}}$ maintains a stable orientation at $(100\pm20)^{\circ}$, whereas $\chi_{\rm{mm}}^{\rm{Qs}}$ is roughly perpendicular to this ($\sim0^{\circ}$), as expected for a plane-perpendicular shock wave propagating to the south towards Qs.
Owing to the large peak value of Qs, $p_{\rm{mm,Qs}}^{\rm{max}}\sim16$\,\%, one cannot explain the orthogonal optical-millimeter polarizations by opacity effects.
Instead, we propose that the optical polarization mainly arises in a conical shock associated with the 7-mm core, while the millimeter-wave polarization results from a propagating shock front associated with Qs. 
We surmise that the moving shock also emits polarized optical radiation, since the optical polarization drops precipitously when the orthogonal polarization of Qs peaks (Fig.~\ref{pol}-\emph{right}).

\subsection{Low Probability of Chance Coincidences}
\label{prob}
The relationship among the $\gamma$-ray, optical, and radio-millimeter flares is supported by probability arguments.
If the flares occur randomly, the probability of observing, at any time, a $\gamma$-ray outburst like the one reported here (i.e., with flux $\gtrsim10^{-6}\,{\rm{phot}}\,{\rm{cm}}^{-2}\,{\rm{s}}^{-1}$ and duration $\sim70$ days) is $p_{\gamma}=0.08$.
For optical and radio--millimeter wavelengths, this probability is $p_{\rm{opt}}=0.04$ and $p_{\rm{mm}}=0.15$, respectively.
Thus, if the flares at different wavelengths were random and independent of each other, the probability of observing a $\gamma$-ray, optical, and radio-millimeter flare at any given time is $p_{\gamma,{\rm{opt}},{\rm{mm}}}=5\times10^{-4}$. This counters the null hypothesis of random coincidence at 99.95\,\% confidence.
This confidence keeps $>99.6$\,\% even when unrealistically considering twice as long as observed $\gamma$-ray, optical, and radio-millimeter flares, thus pointing out the robustness of this result.

\section{Discussion and Conclusions}

The coincidence of the ejection and propagation of Qs---by far the brightest non-core feature ever reported in \object{0235+164}---with the prominent $\gamma$-ray to radio outbursts and the extremely high values of $p_{\rm{opt}}$ and $p_{\rm{mm,Qs}}$ provides convincing evidence that all these events are physically connected. 
This is supported by probability arguments and by our formal DCF analysis, which unambiguously confirms the relation of the $\gamma$-ray outburst in late 2008 with those in the optical, millimeter-wave (including the 7-mm VLBI core) and radio regimes.
Comprehensive studies of other blazars places the core parsecs downstream of the BH \citep{Jorstad:2010p11830,Marscher:2008p15675,Marscher:2010p11374,Agudo:2011p14707}.

\begin{figure*}
   \centering
   \includegraphics[clip,width= 16.cm]{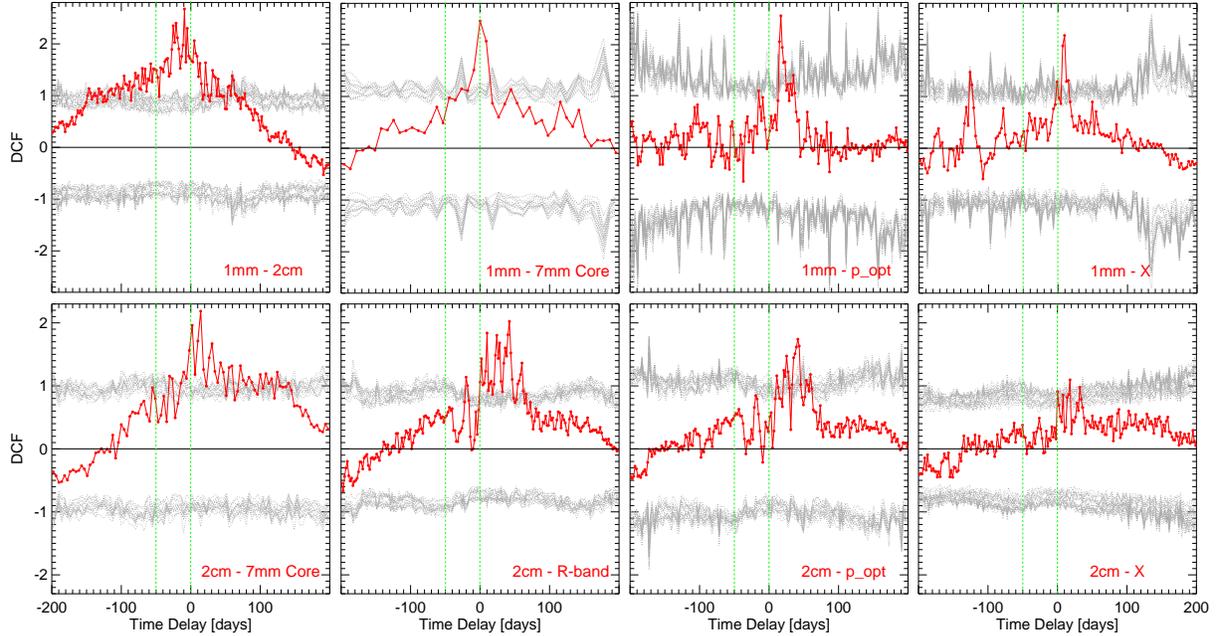}
   \caption{Same as Fig.~\ref{dcf1} but for the 1\,mm light-curve (\emph{top}) and the 2\,cm light-curve (\emph{bottom}).}
   \label{dcf2}
\end{figure*}

The general correlation of $\gamma$-ray and optical light curves (Figs.~\ref{dcf1} and \ref{dcf2}) contrasts with the poor detailed correspondence during the main outburst (Fig.~\ref{tflux}). 
This is difficult to reconcile with the external Compton mechanism, in which $\gamma$-rays arise from inverse Compton scattering of photons originating outside the jet by electrons in the jet \citep{Begelman:1987p15667,Dermer:1993p15668}. 
According to this model, rapid fluctuations of the $\gamma$-ray and optical flux are caused by changes in the number of electrons with sufficient energy to radiate at these frequencies. 
However, such changes should affect the flux at both wavebands in a similar manner if the seed photon field varies smoothly with distance down the jet, as expected if these photons are from the broad emission-line region \citep{Sikora:1994p15669} or a dust torus \citep{Bazejowski:2000p12923,Malmrose:2011p15425}. 
This is contrary to our observations unless the magnetic field varies in a much different manner than does the energy density of relativistic electrons. 
Instead, the nature of the variations suggests that the seed photon field also varies quite rapidly.

A naturally variable seed photon source is synchrotron radiation from the jet. 
The maximum apparent $\gamma$-ray luminosity of \object{0235+164} is similar to that of the optical synchrotron radiation \citep{Abdo:2010p11947}. 
Thus, synchrotron self-Compton (SSC) radiation can potentially explain the $\gamma$-ray emission without leading to higher-order scattering that flattens the $\gamma$-ray spectrum. 
We can reconcile the short time-scales of variability, $\lesssim1$~day, with a distance from the BH of parsecs, by appealing to both the very narrow opening angle of the jet (see above) and the possibility of structure across the jet. 
The former limits the width of the emission region to $\sim0.8$~pc, while the latter allows individual zones of emission to be no larger than $\sim0.8/\sqrt{N}$~pc, where $N$ is the number of zones, which presumably represent turbulent cells. 
In fact, \citet{Marscher:2010p14415} used \object{0235+164} as a blazar whose spectral energy distribution, polarization, and variability properties can be well represented by such a model. 
The synchrotron flux varies in response to changes in the energy density of the flow, maximum electron energy, and magnetic field direction in each cell, while the SSC flux varies from the first two of these plus the time-delayed synchrotron photon density from the other cells. 
These variations follow similar trends, leading to a general correlation but with detailed fluctuations that can differ significantly.

We identify the 7-mm core as the first re-collimation shock near the end of the jet's acceleration and collimation zone \citep[ACZ][]{2007AJ.134.799J,Marscher:2008p15675,Marscher:2010p11374}.
Superluminal feature Qs is consistent with a moving shock oriented transverse to the jet axis, given the extremely high $p_{\rm{mm,Qs}}$, with $\chi_{\rm{mm}}^{\rm{Qs}}$ parallel to the direction of propagation of Qs.
The flux evolution of the core appears closely tied to that of Qs, and its light curve is correlated at high confidence with those at $\gamma$-ray, optical, and millimeter wavelengths.
This suggests that Qs is the head of an extended disturbance, perhaps containing a front-back structure stretched by light-travel delays in the observer's frame \citep[see, e.g.,][]{Aloy:2003p350}.
The presence of this dual structure is also consistent with the perpendicular $\chi_{\rm{opt}}$ and $\chi_{\rm{mm}}^{\rm{Qs}}$ if Qs corresponds to the front region emitting at millimeter wavelengths, while the optical flare arises mainly from the interaction of the slower back region with the standing conical shock in the core.
 
Under this scenario, the radio/millimeter-wave and optical (and perhaps X-ray) synchrotron flares start when the front region crosses the conical shock at the core, where the jet is at least partially optically thin.
This interaction accelerates electrons, increasing the synchrotron emissivity, and produces SSC $\gamma$-ray emission from the up-scattering of IR-optical photons. 
There is a time delay of the latter, since the seed photons must travel across part of the jet before arriving at the scattering site \citep[see][]{Sokolov:2004p12696}. 
When the back region of the moving perturbation encounters the core, their interaction again produces efficient particle acceleration, which is seen as a sudden optical and radio/millimeter synchrotron emission enhancement.
The subsequent optical variability is produced by the passage of the remaining shocked turbulent plasma in the back structure through the core.
During the different optical sub--flares, the integrated radio/millimeter synchrotron flux keeps rising.
This radio/millimeter outburst is more prolonged owing to the longer synchrotron cooling-time of electrons radiating at these wavelengths and the lower speed of the back structure.
Indeed, Qs does not reach its maximum radio/millimeter-wave flux until traveling a projected distance of $\sim0.13$ mas from the core.

When the entire front-back structure passes across the core, the synchrotron emission declines rapidly at optical (and, if relevant, X-ray) frequencies, as does the $\gamma$-ray SSC emission.
The decay of the radio/millimeter-wave emission is more gradual (see above).
The prominent X-ray flare on RJD$\sim$54760--54770 is consistent with being simultaneous with the peak of a millimeter and/or optical synchrotron flare and could arise from either scattering of millimeter photons or synchrotron radiation from the highest-energy electrons.
The former is consistent with the evolution of the X-ray spectral index, which gradually steepens to values consistent with the optically thin millimeter--IR spectral index ($\sim1$) at the time of the X-ray peak.

\begin{acknowledgements}
We acknowledge the anonymous referee for constructive comments.
This research was funded by NASA grants NNX08AJ64G, NNX08AU02G, NNX08AV61G, and NNX08AV65G, NSF grant AST-0907893, and NRAO award GSSP07-0009 (Boston University); RFBR grant 09-02-00092 (St.~Petersburg State University); MICIIN grant AYA2010-14844, and CEIC (Andaluc\'{i}a) grant P09-FQM-4784 (IAA-CSIC); the Academy of Finland (Mets\"{a}hovi); NASA grants NNX08AW56S and NNX09AU10G (Steward Observatory); and GNSF grant ST08/4-404 (Abastunami Observatory).
The VLBA is an instrument of the NRAO, a facility of the NSF under cooperative agreement by AUI. 
The PRISM camera was developed by Janes et~al., and funded by NSF, Boston University, and Lowell Observatory. 
Calar Alto Observatory is operated by MPIA and IAA-CSIC. 
The IRAM 30m Telescope is supported by INSU/CNRS, MPG, and IGN.
The SMA is a joint project between the SAO and the Academia Sinica. 
\end{acknowledgements}


\end{document}